\documentclass[sigconf,screen,authorversion]{acmart} 

\usepackage[utf8]{inputenc} 		
\usepackage{csquotes}				
\usepackage{balance}				
\usepackage[nameinlink]{cleveref}	
\usepackage[subtle]{savetrees}		

\copyrightyear{2019} 
\acmYear{2019} 
\setcopyright{acmlicensed}
\acmConference[SIGCSE '19]{Proceedings of the 50th ACM Technical Symposium on Computer Science Education}{February 27--March 2, 2019}{Minneapolis, MN, USA}
\acmBooktitle{Proceedings of the 50th ACM Technical Symposium on Computer Science Education (SIGCSE '19), February 27--March 2, 2019, Minneapolis, MN, USA}
\acmPrice{15.00}
\acmDOI{10.1145/3287324.3287448}
\acmISBN{978-1-4503-5890-3/19/02}

\renewenvironment{quote}{\list{}{\leftmargin=1.3em\rightmargin=1.3em}\item\relax\it\hspace{-0.2em}\textquotedblleft\ignorespaces}{\unskip\unskip\textquotedblright\endlist}

\definecolor{darkgreen}{rgb}{0.15, 0.57, 0.1}
\definecolor{darkblue}{rgb}{0.15, 0.1, 0.57}
\definecolor{darkred}{rgb}{0.31, 0, 0.}

\begin{document}

\title{Reflective Diary for Professional Development of~Novice~Teachers}

\author{Martin Ukrop} 
\orcid{0000-0001-8110-8926}
\affiliation{
  \institution{Masaryk University}
  \country{Czech Republic}
}
\email{mukrop@mail.muni.cz}

\author{Valdemar Švábenský} 
\orcid{0000-0001-8546-280X}
\affiliation{
  \institution{Masaryk University}
  \country{Czech Republic}
}
\email{svabensky@ics.muni.cz}

\author{Jan Nehyba} 
\orcid{0000-0003-4159-5576}
\affiliation{
  \institution{Masaryk University}
  \country{Czech Republic}
}
\email{nehyba@ped.muni.cz}

\begin{abstract}
Many starting teachers of computer science have great professional skill but often lack pedagogical training. Since providing expert mentorship directly during their lessons would be quite costly, institutions usually offer separate teacher training sessions for novice instructors. However, the reflection on teaching performed with a significant delay after the taught lesson limits the possible impact on teachers. To bridge this gap, we introduced a weekly semi-structured reflective practice to supplement the teacher training sessions at our faculty. We created a paper diary that guides the starting teachers through the process of reflection. Over the course of the semester, the diary poses questions of increasing complexity while also functioning as a reference to the topics covered in teacher training. Piloting the diary on a group of 25 novice teaching assistants resulted in overwhelmingly positive responses and provided the teacher training sessions with valuable input for discussion. The diary also turned out to be applicable in a broader context: it was appreciated and used by several experienced university teachers from multiple faculties and even some high-school teachers. The diary is freely available online, including source and print versions.
\end{abstract}


\maketitle

\section{Introduction}

Many computer science (CS) faculties employ teaching assistants (TAs) from the ranks of students. This is done for multiple reasons: to integrate prospective Ph.D. and graduate students into academia, to offload the core employees or to cope with the increasing student enrollment~\cite{Forbes,Vihavainen}.

Multiple CS institutions report overwhelmingly positive experience with student TAs~\cite{Dickson,Roberts,Reges}. They are proactive, enthusiastic and often able to explain complicated concepts to their fellow students well. Since they still remember the student struggles and can adjust the explanations accordingly, they de facto provide a form of peer tutoring~\citep{beasley}, which is often beneficial~\citep{topping}. Moreover, the TAs themselves strongly benefit from the experience they acquire through teaching. Regardless of whether teaching is their long-term career goal or not, they improve their understanding of CS and soft skills.

\begin{figure}[t]
\includegraphics[width=\columnwidth]{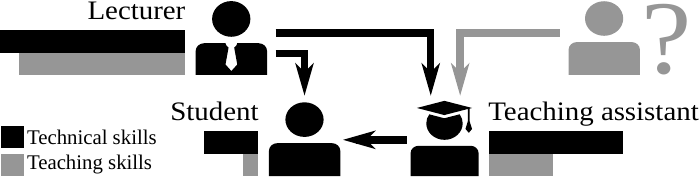}
\caption{The lecturer oversees the technical development of TAs. But who is helping TAs improve their teaching skills?}
\label{fig:problem}
\end{figure}

The TAs usually have excellent academic performance. However, being knowledgeable in CS does not automatically equal the ability to teach it, and the TAs often have little or no prior pedagogical training\footnote{The TAs in other STEM fields are also poorly prepared for their teaching duties~\cite{dechenne}.}. Although the lecturer oversees the development of their technical skills, there is usually nobody helping them improve their pedagogical skills (see \cref{fig:problem}). Instead, they are often expected to acquire teaching skills \enquote{on the fly}, which decreases the positive impact they can have as teachers and often also lowers their self-esteem. Therefore, the need for TA training is universal, and there were efforts to address it in the context of CS~\cite{Estrada,Patitsas}.

In 2016, we established a teacher training course for student TAs at our faculty called the \enquote{Teaching Lab}. We aim to enhance the quality of teaching CS by improving the skills of new TAs. We also cultivate an active community of CS teachers who share their experiences and best practices. While many interventions can improve teaching, we chose to focus on TAs since they form an important (and often overlooked) portion of the university teaching process.\footnote{For the rest of the paper, we will use the term \textit{mentor} as a teacher of the teacher training session, \textit{TA/teacher} as an attendee of the teacher training session, and \textit{student} as a learner who is taught by the TAs.}

Apart from several improvements, the Teaching Lab also introduced two major challenges. Firstly, as literature reviews suggest, there is little evidence regarding the impact of training on teaching~\cite{Gilbert}. Secondly, the effective improvement of university teachers is based on individual feedback, discussion, and mentoring~\cite{Gibbs2004}, but there are not enough skilled teachers willing to provide such expert feedback.

We decided to alleviate both these issues by guiding the teachers to self-reflection and subsequent analysis of their experience in the community. By doing so, we aim to enhance their teaching skills, since \enquote{effective reflective practice is [\dots] a beginning point in the development of professional knowledge}~\cite[p. 38]{loug}.

To support this self-reflection, we created a reflective diary: a semi-structured tool guiding the novice teachers to awareness of their acts (and effects thereof), possible improvements and their strengths/weaknesses. The reflection performed at (or shortly after) the lesson is then revisited and discussed at the regular meeting with other teachers and mentors. The diary also includes a reference to key concepts discussed during the teacher training sessions. Both these aspects allow the diary to bridge the gap between the teacher training sessions and the actual teaching practice of novice TAs.

This paper reports our experience with using the diary and the lessons we have learned. There are two substantial novelties in our approach. Firstly, to the best of our knowledge, the use of reflective diaries to improve the teaching skills of university TAs has not been published so far. Secondly, the diaries used to support reflection are usually much less structured and lack the methodical section. The diary is publicly available online under a Creative commons license~\cite{diary}, and we encourage all fellow teachers from other institutions to use it or adapt it.

\section{Related Work}
\label{sec:related-work}
This section deals with reflective practice and its broader context in CS. In particular, it focuses on multiple versions of reflective diaries and their efficiency.

\subsection{Reflective Practice in Teaching}

One of the first formalizations of reflective practice can be traced to Dewey's concept of reflective thinking~\cite{dewey}. In the second half of the 20\textsuperscript{th} century, reflective practice was systematically developed in the works of Schön~\cite{schon}, Kolb~\cite{kolb}, Boud~\cite{boud} and Brookfield~\cite{brookfield}. Although there is no single accepted definition of reflective practice~\cite{pedro,Hatton}, a common one states it is \enquote{a process of learning from experience towards gaining new insights of self and/or practice for future act}~\cite[p. 15]{finl}. Intuitively, reflective practice is systematic thinking about one's own experience focused on future improvement.

Reflective practice is a core concept for educators and one of the major trends and promising innovations in higher education~\citep{Tembri}. Reflective practice as a strategy for professional development is applied in many contexts, such as pre-service teaching, nursing education, medical education, psychotherapy training, management education or engineering education (cf.~\cite{bassot}). Furthermore, the concept of reflective practice is connected with many other educational approaches, such as self-regulated learning~\cite{wallin}, experiential learning~\cite{harvey} or work-based learning~\cite{siebert}. In these approaches, many methods for supporting professional development are used, including reflective diaries, lesson reports, surveys and questionnaires, audio and video recordings or observation and action research~\citep{Fatemi}.

Reflective writing has been successfully used also in the area of CS education: reflective activities help improve software design using UML~\cite{coffey} and blogging is utilized as a tool for self-directed learning~\cite{mcdermott} or as a form of feedback for course evaluation~\cite{stone}. Furthermore, reflective writing is used for the continuous education of primary and secondary school CS teachers~\cite{reding}. However, reflective diaries as described above mostly focus on supporting the learning of CS \textit{students}. Our approach, on the other hand, innovatively concentrates on novice university \textit{teachers}. Although merely \enquote*{thinking about how you teach} may seem obvious, proper reflective practice is hard to carry out and even harder to teach~\cite{finl}. Therefore, we provide a tool -- the reflective diary -- to help with performing the reflection. 

\subsection{The Variety of Reflective Diaries}
\label{subsec:related-work-diaries}

The reflective diary (a.k.a. learning/teaching journal) is one of the most prominent tools for reflective practice. It is a \enquote{container for writing that provides students with a framework to structure their thoughts and reflections}~\cite[p. 6]{wallin}. This learner-centered activity provides opportunities for self-education and improves performance.

Although the literature contains general recommendations for reflective diaries (regarding their content, structure or activities performed with it)~\cite{moon}, usually only ad-hoc supplementary materials are available~\cite{diary1,diary2,diary3} (and often not the complete diaries).

There are many different types of reflective diaries. We can divide them according to three basic criteria~\cite{moon}: the internal structure, the number of co-authors and the medium they use.

\subsubsection{Internal Structure}
Different elements can provide the structure within the diary. Examples include a set of standard or contextualized questions, a timeline for a critical incident, comments written or drawn by the learner or a box for an external supervisor (for more examples, see~\cite{trif}). A well-structured reflective diary can function as a formative assessment of the performer (thus helping the teacher improve their teaching skills). For novice teachers, a more structured reflective diary seems to work better than a less structured one~\cite{bassot2015}.

\subsubsection{Number of Authors}
A single-author diary serves as a tool for personal development using only self-exploration. A dialogical diary, on the other hand, allows cooperation between the primary author and their supervisor. The access of the supervisor may be limited in some way -- e.g., only to observe or to comment on existing sections. A particular kind of dialogical diary is a fully collaborative journal by multiple authors.

\subsubsection{Medium Type}
A reflective diary can be either paper-based or digital. Although a printed version can be seen as a bit more personal, the digital version is more versatile. Examples include blogs used as personal diaries or to retain a record of events, audio recordings (monologues or dialogues), videos or wiki-based systems for collaborative reflective practice.

\subsection{The Evaluation of Reflective Diaries}
Research shows that a reflective diary is a useful tool for improving both student and teacher performance. On the one hand, literature reviews document the effectiveness of diaries for university \textit{students}, enabling them to advance in their field of study -- examples include pre-service teachers~\cite{lindroth}, nursing teachers~\cite{epp} and higher education in general~\cite{dyment}. On the other hand, a reflective diary was used for supporting the career of starting university \textit{teachers}~\cite{bell}. In an open-ended questionnaire of 64 participants, 80\% found journal writing beneficial. According to their opinions, it helped to reflect on their teaching (80\%), improved their teaching (81\%), and helped link theory and practice (67\%)~\cite{bell}. Our paper is of more qualitative nature, focusing more on the experience from using the reflective journal.

\section{The Reflective Diary}
\label{sec:diary}

This section presents the reflective diary as a tool for integrating teacher training sessions with teaching practice. Firstly, we explain the context leading to the creation of the diary and the principles we employed. Secondly, we outline individual sections of the diary and discuss their aim. Lastly, we give details on a pilot use of the diary that provided us with the lessons learned described in \cref{sec:lessons-learned}.

\subsection{The Wider Context}
\label{subsec:context}

The diary was created to improve teacher training sessions at the Faculty of Informatics, Masaryk University, Czech Republic. It is a CS faculty with approximately 1,400 students and about 460 people participating in teaching. Out of these, roughly 100 is the regular faculty staff, 80 are members of other faculties and institutions at the university, 110 are external contractors, 55 are Ph.D. students, 50 are graduate (Master's) students, and the remaining 65 are undergraduate (Bachelor's) students. Every semester, about 40 new people start teaching (mostly bright graduates or beginning Ph.D. students who start as TAs). At the same time, a similar number of people stop teaching (mostly graduate students leaving the faculty), keeping the total number somewhat steady. 

The TAs participate in a wide variety of courses. Content-wise, the courses span a full CS curriculum, including programming, software engineering, algorithms, formal languages, and discrete mathematics. The form of the courses also varies considerably: from courses featuring explanation in front of the board to those involving individual computer work. As a result, the roles of the TAs vary across the courses: they can lecture, tutor, mentor, grade assignments or consult projects.

As can be seen, students at our faculty form an essential part of teaching. Thus, there is a crucial need to develop their skills. In the past, however, the faculty did not offer any pedagogical training to the student TAs. This situation changed in 2016 when we introduced the Teaching Lab (a teacher training course) and started building the community of skilled teachers open to discussing the issues they face in teaching. The course is intended for approximately 20 attendees who are actively tutoring seminars. It is not mandatory, but 3 credits are awarded for successful accomplishment. In the weekly sessions, we introduce TAs to a variety of relevant topics, including but not limited to: group interactions, asking clear questions, using common teaching tools, creating innovative activities, lesson planning and defining learning outcomes.

We highlight that Teaching Lab is a bottom-up initiative of the TAs from inside the faculty. As a result, the teacher training is not generic but takes into account the specifics of teaching CS. Moreover, it encourages discussion on how to improve particular courses at our institution. So far, the course opened three times.\footnote{In spring 2016, fall 2016 and fall 2017. We are preparing a fourth run for fall 2018.} In the latest run, we introduced the reflective diary.

\subsection{Creating the Reflective Diary}
\label{subsec:diary-creating}
Our motivation for creating the reflective diary was twofold. Firstly, we wanted the TAs to become reflective practitioners in teaching. Since literature suggests that teachers do not often reflect on their teaching~\cite{Heng}, we are focusing (in accord with Hussein~\cite{hussein}) on the means of supporting reflection. The diary can function as the first step to learn from own progress without the costly need of expert mentors. Secondly, we wanted to encourage using the content (such as concepts and tools) discussed in the Teaching Lab classes. If novice teachers apply the discussed concepts and tools themselves, they gain personal experience on what works and when.

When developing the diary, we had two additional design goals: to keep it modular (and thus widely applicable) and easily approachable even by unskilled teachers. The high modularity is necessary because of the heterogeneity of courses taught by our novice teachers. Furthermore, it may allow the adoption of the diary outside of the Teaching Lab community. The reason for easy approachability is that the primary target audience is undergraduate and graduate students teaching their first semester, often with no prior formal teacher training.

Keeping the motivations and design goals in mind, a handful of active and experienced students and teachers (mainly the authors of the paper) drafted the structure of the diary. In comparison with the diaries listed in \cref{subsec:related-work-diaries}, ours is more structured and enriched with the methodical materials. The process continued by gathering and incorporating suggestions from other members of the Teaching Lab community. It took about three months to develop the first release version from the conception of the idea to print. We preferred paper as a medium as it is easier to develop and work with compared to electronic applications.

\subsection{Structure of the Diary}
\label{subsec:diary-structure}

\begin{figure*}
\includegraphics[width=\textwidth]{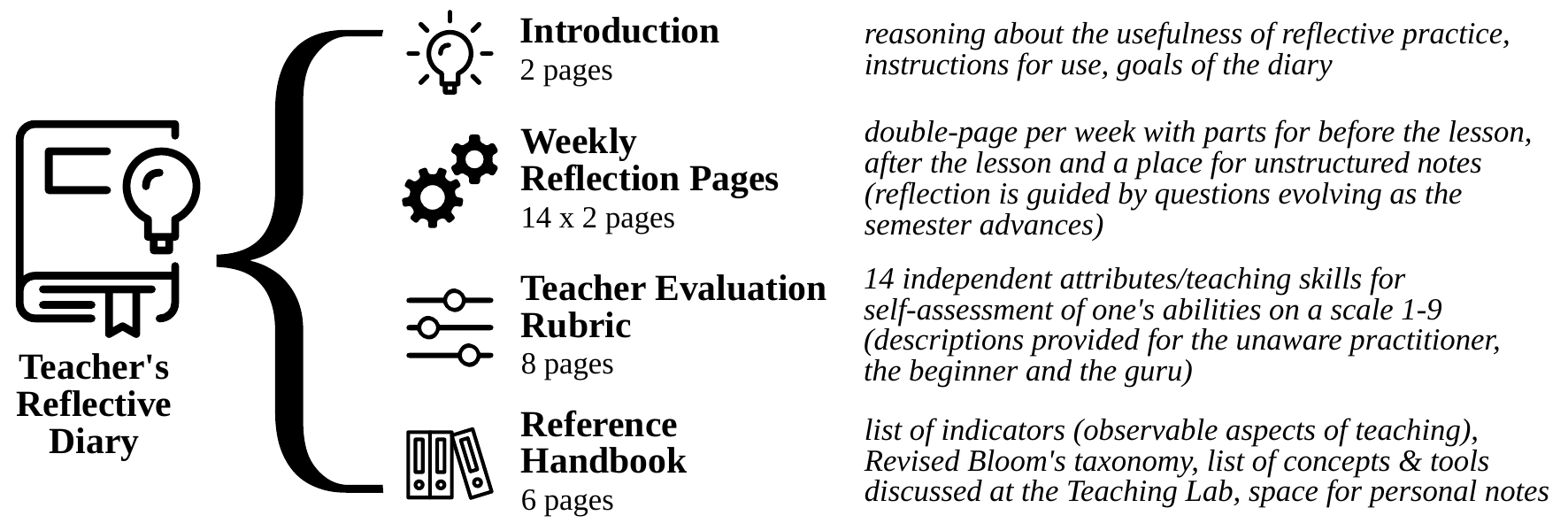}
\caption{An overview of the Teacher's reflective diary and content of the individual sections. The diary is freely available on GitHub~\cite{diary} under a Creative Commons license.}
\label{fig:diary}
\end{figure*}

A basic overview of the reflective diary is shown in \cref{fig:diary}. It has 48 pages in A6 format to comfortably fit into one's pocket. After opening the diary, the introduction explains the importance of reflection. It is followed by the structured space for individual weeks guiding the TA's regular reflection. The diary concludes with a teacher evaluation rubric reminding you what you want to improve and a short reference handbook for concepts and tools discussed in the Teaching Lab sessions. The diary is typeset in \LaTeX{} and version-controlled, allowing for easy modification. It exists in English and Czech versions. Both the sources and the compiled documents ready for print are available on GitHub~\cite{diary}.

\subsubsection{Diary Introduction}
The diary starts with brief reasoning about the usefulness of reflective practice in teaching and several instructions for use (most notably to use it regularly). To aid the correct use, the main goals of the diary are then transparently summarized: 1) it reminds one to reflect on one's teaching, 2) it provides a convenient place to collect notes for the future, 3) it helps to see different aspects of teaching and 4) it enables to track progress.

\subsubsection{Weekly Reflection Pages}
The core of the diary is formed by 14 double-page spreads providing enough reflection space for all semester weeks. Each consists of a part to be filled in before the lesson, another part for after the lesson and a full page for notes. The \enquote*{before} part encourages the teacher to plan the timeline of the lesson, set learning outcomes or prepare other interaction (e.g., \textit{What precedents do I want to set?}, \textit{What questions do I plan to ask the group?}). The \enquote*{after} part serves for recording subjective satisfaction (\textit{What is my lesson satisfaction?}), highlighting the most prominent successes and opportunities for improvement (\textit{What worked well? What could have been better?}) as well as other useful attributes (e.g., \textit{Which questions did work in the class? Which did not?}). The extra space for comments always has multiple questions pre-printed in gray to guide the teacher if they cannot think of anything to note down (e.g., \textit{How many group questions did I ask? How many questions did the students ask me? Was that too few, enough or too many?}).

The structure and questions on the weekly reflection pages evolve throughout the semester to gradually deepen the reflection. In the beginning, the questions are accompanied by short explanations (e.g., \textit{What is the structure of the lesson? Outline 2–6 blocks on the timeline below.}). Even more importantly, the questions change according to what is relevant in the particular part of the semester: At the beginning, they revolve around precedents, norms, (un)wanted interactions, asking the right questions or identifying teaching problems. In the end of the semester (as more advanced concepts are discussed at the Teaching Lab sessions), they concern higher-level thinking such as what does the course final exam test, whether the course has adequately prepared the students for the exam, how to collect feedback on teaching or what progress did the teacher achieve during the semester.

\subsubsection{Teacher Evaluation Rubric}
The next part is a scoring rubric for self-assessment of teaching skills, which reminds the teachers of what they want to improve. It consists of 14 independent attributes/teaching skills (not exhaustive but quite representative). For each attribute, one can self-assess their skills on a scale from 1 to 9 with three prominent points described in detail (the unaware, the beginner and the guru). We selected the skills based on what we focus on in the Teaching Lab, which also overlaps with best teaching practices in the literature~\cite{petty,Leyzberg}.

Let us illustrate it on an example of a single attribute -- \textit{Flexibility/adjustments on the spot}. The label of the teacher unaware of this aspect of teaching reads: \textit{Unaware: I do not consciously respond to situations arising at the lesson.} The beginner practitioner is described as: \textit{Beginner: I am aware of moments where it may be interesting or useful to deviate from the intended lesson plan. Nevertheless, I’m usually unable to react appropriately on the spot.} and the guru level is described as: \textit{Guru: I’m able to adjust my lessons on the fly according to the situation and students’ needs. I know enough tools and can use them effectively.}

The usage of the scoring rubric is twofold. Primarily, it allows identifying one's weak spots and tracking improvement. At the beginning of the semester, the owner of the diary assesses themselves in all areas in the rubric and chooses 1--3 skills to concentrate on in the upcoming semester. They can also think of specific indicators if they want to be extra thorough. At the end of the semester, they complete the whole rubric again to see their (self-perceived) progress. Another option is to treat the rubric as a manifesto -- the guru descriptions represent our view of the skills great teachers have. Naming and listing these skills enables the teachers to reflect on them with their colleagues.

\subsubsection{Reference Handbook}
The last part of the diary is a small handbook, which summarizes the concepts and tools discussed during the teacher training sessions. There is a page on indicators (directly observable aspects of teaching) that the teacher can track to help them reflect, evaluate and improve their lessons. Next, the Revised Bloom's taxonomy~\cite{bloom} is summarized along with suggested action verbs to help the teacher think about learning outcomes. Then the diary includes a list of teaching tools introduced at the Teaching Lab sessions. This list eases the adoption of the tools by novice teachers as it works as a short and comprehensive reminder. The tools span across multiple aspects, from the ones that help structure the lesson better, through ones that help assign tasks clearly and efficiently, to ways to create new tasks/exercises or to think about the broader context of the course. The section ends with several empty pages for the owner's personal remarks.

\subsection{The Pilot Use of the Diary}
\label{subsec:diary-use}
In the latest run of the Teaching Lab in fall 2017, we deliberately focused on reflective practice. Before the start of the semester, we handed out reflective diaries to 25 attendees of the course as a compulsory part of the training. Each week, they journaled about their teaching experiences, reflected on them, and shared their findings during the Teaching Lab sessions with their peers and the mentors. Furthermore, we complemented the self-reflection with compulsory peer evaluation, where they observed a peer's teaching practice and provided feedback. Additionally, we offered the diary to several interested senior teachers, who volunteered to try it out. At the end of the semester, we collected feedback from all the participants in an audio-recorded focus group discussion and over e-mail.

Our case study has some limitations. First, there is a strong self-selection bias regarding the users of the diary. Since Teaching Lab is an optional course, its attendees (and thus the users of the diary) are interested in learning how to teach better. The selected staff members intrigued to use the diary also tend to invest an above-average time and effort to perfect their courses. Second, the results may be further biased since we gathered feedback only from a single institution (and only from starting CS teachers and a few language teachers). Nevertheless, the diary was primarily meant to help novice CS teachers, so inapplicability in other fields is not that troublesome. Third, an interviewer bias might be present as the diary authors were present in the focus group discussion and also evaluated the e-mail feedback. To mitigate this, three people analyzed the inputs independently.

\section{Lessons Learned}
\label{sec:lessons-learned}

This section presents the lessons learned based on the experience of about 30 teachers from our faculty who used the diary (25 student TAs attending the Teaching Lab and about five members of the regular staff). We share the gathered feedback (including anonymized direct quotations) complemented by our own experience. Furthermore, we discuss the implications and, based on them, suggest future improvements in the next section.

\subsection{Successes}
\label{subsec:successes}
We present three significant successes: Using the diary supported the teachers in reflecting on their practice and connected the teacher training sessions to teaching practice more tightly. Moreover, the diary had a broader impact extending outside of our faculty.

\subsubsection{Supporting the Reflective Practice}
Owning a diary worked as a reminder for the teachers to reflect on their experiences after each lesson. We hypothesize that the diary not only supported but often initiated the process of reflection.
\begin{quote}
To a great extent, the diary worked as a \enquote*{kick} to sit down and think about my lesson.
\end{quote}
\begin{quote}
Without the Teaching Lab and the diary, I would not think this thoroughly about how I teach.
\end{quote}
Perhaps surprisingly, for some, the diary also worked as a motivator to teach better.
\begin{quote}
[The diary] had a motivational effect: It made me plan something remarkable so that I would be able to write it into the section listing what went well.
\end{quote}
Next, the diary allowed teachers to gather ideas before, during, and after a lesson. It helped them to keep track of the notes and suggestions for future improvements.
\begin{quote}
I could tidy up all the ideas of what went well and what didn't in a single place. If the yellow diary didn't shine brightly on my desk, I would probably not do it at all. And since it's in the diary, I know where to find it [...]
\end{quote}
\begin{quote}
[I will use it] even more the following year when I'll be checking before every lesson to see what didn't work and what I did wrong.
\end{quote}
Moreover, using the diary raised awareness about reflecting on one's teaching, which cultivated the environment at the faculty. By journaling, the TAs formulated ideas for discussion among themselves. In hindsight, they considered using the diary to be a good experience and did not oppose to having it compulsory.

\subsubsection{Connecting Teacher Training Sessions and Actual Teaching}
The diary also acts as a handbook: it summarizes concepts and tools discussed in the Teaching Lab sessions. This helped the starting teachers try out and perfect new approaches in their classes.
\begin{quote}
[Thanks to the diary] I have a list of tools and activities to use together in one place.
\end{quote}
What is more, the diary seems to have increased the teachers' preparation effort, which, arguably, improved their teaching.
\begin{quote}
[Thanks to the diary] I had put more effort into lesson preparation during the semester. I really filled it in up to about 11\textsuperscript{th} week of the semester.
\end{quote}
\begin{quote}
[...] before the lesson I noted down \enquote*{this must be mentioned to students} so as not to forget.
\end{quote}

\subsubsection{Having a Broader Impact}
As it was initially not intended, we consider reaching people outside the Teaching Lab community as a major success. The diary attracted the attention of several other student TAs and a few employees of other institutions within our university. Some of them actively used the diary and provided feedback corresponding to the findings above. Furthermore, when we mentioned the diary at relevant events (such as ACM ITiCSE), multiple individuals intended to adapt and use it and promised to share their experience. Teacher training centers at three major Czech universities consider adopting the diary as well. Outside the university realm, a few secondary school teachers were excited and used the diary.

\subsection{Challenges}
\label{subsec:challenges}
The adopters of the diary encountered three major challenges: irregular use, clashes with other tools and a poor fit for the class format.

\subsubsection{Dropout and Irregular Use}
As could be expected, not everyone used the diary regularly for the whole semester. Especially people not enrolled in the Teaching Lab (i.e., those not required to use it) dropped out after several weeks.
\begin{quote}
I wanted to use it, I printed it, but my enthusiasm lasted only for the first few weeks (about four).
\end{quote}
Others did not use it soon enough after their lesson. This is a problem, since those who did not reflect on a lesson immediately or shortly after usually forgot the details or did not reflect at all. 
\begin{quote}
I have a bad experience with filling the diary in too late [after the lesson] [...] I appeal to everyone to reserve 5--10 minutes right after the lesson [to fill in the diary].
\end{quote}
What is more, the part \enquote*{before the lesson} should be completed before a lesson.
\begin{quote}
[I would advise] to fill the \enquote*{before} part really before the lesson. In 75\% of the cases, I filled it in only after the lesson -- but in the remaining 25\% it really helped me to think more about the upcoming lesson.
\end{quote}

\subsubsection{Clashes with Other Tools}
Some teachers (both those enrolled in the Teaching Lab and senior ones) already used other means of lesson planning and evaluation (paper or electronic). Therefore, updating the diary sometimes clashed with other tools.
For example, some courses require TAs to write weekly reports from the seminars to the leading lecturer. In these cases, people were often unsure of what to write in the diary and what to write in the report.
\begin{quote}
[...] when we write it [in the report] to the professor, one does not want to rewrite everything into the diary.
\end{quote}
Other courses feature errata documents for the course materials (such as slides or exercises). When a teacher discovers an error and notes it down in the diary (which is conveniently accessible when reflecting), they have to rewrite it elsewhere later on.

\subsubsection{Poor Fit for Class Format}
The reflective diary was created mainly with weekly lessons in mind. Therefore, the structure did not fit some TAs teaching other types of classes.
\begin{quote}
I taught two seminar groups -- each one bi-weekly and thus the seminars repeated. Apart from that, I taught no theory [...] if I taught [weekly] \enquote*{Mathematical Foundations of Computer Science}, I would use the diary more often or more effectively.
\end{quote}
As mentioned in \cref{subsec:context}, the teaching approaches and formats used in different courses vary significantly. The above observation shows that the diary is not yet modular enough to accommodate this.

\subsection{Observations}
\label{subsec:observations}
Finally, we share participants' comments that can be classified as neither positive nor negative. These include comments on the medium, format and amount of internal structure. Overall, the diary sections are structured loosely enough to be easily adaptable to one's needs. However, this might have resulted in different ways of use and thus different (and sometimes contradicting) wishes.

\subsubsection{Medium and Format}
Some TAs suggested that an electronic version (e.g., a mobile app), would be more practical. Others opposed this view and liked the printed version.
Of these, some would appreciate a bigger format (e.g., A5-sized) to accommodate more detailed lesson plans while others liked the small and handy A6 size.

\subsubsection{Amount of Internal Structure}
Finally, some users (especially the beginners) wanted even more questions in the section for free notes to guide their reflection. These generally also wanted more content in the handbook section (e.g., extended and more detailed accounts of tools and concepts discussed at the Teaching Lab).
Others (usually the more advanced teachers) sometimes disregarded the guiding questions or had already asked them implicitly and thus used the handbook section minimally.

\section{Discussion and Future Work}
\label{sec:future-work}

We devise directions for future work from the lessons learned described in the previous section. Considering the successes (\cref{subsec:successes}), we plan to keep the reflective diary a compulsory part of the Teaching Lab, since the usage of the diary turned out to be beneficial according to both the mentors and the teachers. Moreover, we will strengthen guided reflection and experience sharing during the sessions to enhance the connection between teacher training and practice. Finally, as we have seen interest from other groups (language teachers, other institutions), it may be worth initiating proper cooperation with them.

Regarding the challenges (\cref{subsec:challenges}), it seems that the dropout is mostly the result of teachers having too much other work and therefore finding the diary expendable. It may be caused by some TAs not perceiving the diary as something that immediately improves their teaching. In the next semester, we will, therefore, emphasize that the benefits of using the diary may become apparent only in the long-term perspective, urging the users to attempt using for a longer time. Furthermore, we will advise them to reserve a short time slot right after the lesson for reflection as doing it later turned out to have limited impact. As for the tool clash (having/using other tools for feedback and reflection), we will ask the teachers what tools they already use and openly discuss how to achieve synergy by incorporating the reflective diary along with them. We will also proactively address how to adjust the individual parts of the diary for teachers with non-standard lessons (e.g., those having only consultations).

As for the observations (\cref{subsec:observations}), we are currently not planning to design an electronic version. Although it could be useful if we wanted to analyze the content of the diaries in more detail, this is not our primary aim, and the development would be quite time-consuming. However, creating two formats (A6 and A5) is feasible and could be helpful for some. Another alternative is a diary with addable/removable pages. Such an adjustable design could optionally accommodate lesson plans or a more detailed version for novices while allowing the advanced teachers to remove the content they would not use anymore.

\section{Conclusion}
\label{sec:conclusion}

We created the Teacher's reflective diary for novice university instructors -- a simple yet powerful tool for professional development of CS teachers. Through guided self-reflective journaling combined with a reference handbook, the TAs practically applied concepts from the teacher training and increased self-awareness of their teaching. Since the TAs used the diary in a wide range of CS courses, we gathered rich insight on what worked well and what could be improved. Moreover, the inputs from the TAs' diaries contributed to an open discussion and sharing of experiences in Teaching Lab sessions, cultivating the teaching environment at our institution.

The diary is unique in providing a structured guide for reflection. Based on the thorough literature review, we argue that the diary is the first tool of its kind focused on TAs in CS. Therefore, we encourage fellow CS teachers to use the diary at their institutions and recommend it to TAs. Both the \LaTeX{} sources and the compiled documents ready for printing are available on GitHub~\cite{diary} for free under a Creative Commons license. While our initial goal was to create the diary for CS university teachers, with very little modification it turned out to work well also outside the context of CS and outside universities.

\begin{acks}
We thank Ondr\'{a}\v{s} P\v{r}ibyla for his contributions to the diary and the paper, Martin Mac\'{a}k and the Teaching Lab community for inputs regarding the diary, Bill Siever, Lydia Kraus and Petr \v{S}venda for a thorough internal review and the Faculty of Informatics, Masaryk University for long-term support of the Teaching Lab initiative. Martin Ukrop was also supported by Red Hat Czech.
\end{acks}

\balance
\bibliographystyle{ACM-Reference-Format}
\bibliography{references}


\begin{thebibliography}{47}


\ifx \showCODEN    \undefined \def \showCODEN     #1{\unskip}     \fi
\ifx \showDOI      \undefined \def \showDOI       #1{#1}\fi
\ifx \showISBNx    \undefined \def \showISBNx     #1{\unskip}     \fi
\ifx \showISBNxiii \undefined \def \showISBNxiii  #1{\unskip}     \fi
\ifx \showISSN     \undefined \def \showISSN      #1{\unskip}     \fi
\ifx \showLCCN     \undefined \def \showLCCN      #1{\unskip}     \fi
\ifx \shownote     \undefined \def \shownote      #1{#1}          \fi
\ifx \showarticletitle \undefined \def \showarticletitle #1{#1}   \fi
\ifx \showURL      \undefined \def \showURL       {\relax}        \fi
\providecommand\bibfield[2]{#2}
\providecommand\bibinfo[2]{#2}
\providecommand\natexlab[1]{#1}
\providecommand\showeprint[2][]{arXiv:#2}

\bibitem[\protect\citeauthoryear{{Alan Chapman}}{{Alan Chapman}}{2006}]%
        {diary2}
\bibfield{author}{\bibinfo{person}{{Alan Chapman}}.}
  \bibinfo{year}{2006}\natexlab{}.
\newblock \bibinfo{title}{{Reflective diary/journal process and notes}}.
\newblock
\newblock
\urldef\tempurl%
\url{https://www.businessballs.com/freepdfmaterials/reflective_diary_journal_templates.pdf}
\showURL{%
\tempurl}


\bibitem[\protect\citeauthoryear{Bassot}{Bassot}{2012}]%
        {bassot}
\bibfield{author}{\bibinfo{person}{Barbara Bassot}.}
  \bibinfo{year}{2012}\natexlab{}.
\newblock \bibinfo{booktitle}{\emph{The reflective diary: enhancing
  professional development}}.
\newblock \bibinfo{publisher}{Matador}, \bibinfo{address}{Leicestershire, UK}.
\newblock


\bibitem[\protect\citeauthoryear{Bassot}{Bassot}{2015}]%
        {bassot2015}
\bibfield{author}{\bibinfo{person}{Barbara Bassot}.}
  \bibinfo{year}{2015}\natexlab{}.
\newblock \bibinfo{booktitle}{\emph{{The reflective practice guide: An
  interdisciplinary approach to critical reflection}}}.
\newblock \bibinfo{publisher}{Routledge}, \bibinfo{address}{London}.
\newblock
\urldef\tempurl%
\url{https://doi.org/10.4324/9781315768298}
\showDOI{\tempurl}


\bibitem[\protect\citeauthoryear{Beasley}{Beasley}{1997}]%
        {beasley}
\bibfield{author}{\bibinfo{person}{Colin Beasley}.}
  \bibinfo{year}{1997}\natexlab{}.
\newblock \showarticletitle{Students as teachers: The benefits of peer
  tutoring}. In \bibinfo{booktitle}{\emph{Proceedings of the 6th Annual
  Teaching Learning Forum}}. \bibinfo{publisher}{Murdoch University},
  \bibinfo{address}{Perth}, \bibinfo{pages}{21--30}.
\newblock


\bibitem[\protect\citeauthoryear{Bell and Gillett}{Bell and Gillett}{1996}]%
        {bell}
\bibfield{author}{\bibinfo{person}{Maureen Bell} {and} \bibinfo{person}{Max
  Gillett}.} \bibinfo{year}{1996}\natexlab{}.
\newblock \showarticletitle{Developing reflective practice in the education of
  university teachers}. In \bibinfo{booktitle}{\emph{Proceedings of the 1996
  Annual Conference of the Higher Education and Research Development Society of
  Australaia}}. HERDSA, \bibinfo{publisher}{Higher Education and Research
  Development Society of Australaia}, \bibinfo{address}{Australia},
  \bibinfo{pages}{46--52}.
\newblock


\bibitem[\protect\citeauthoryear{Boud, Keogh, and Walker}{Boud
  et~al\mbox{.}}{1985}]%
        {boud}
\bibfield{author}{\bibinfo{person}{David Boud}, \bibinfo{person}{Rosemary
  Keogh}, {and} \bibinfo{person}{David Walker}.}
  \bibinfo{year}{1985}\natexlab{}.
\newblock \bibinfo{booktitle}{\emph{{Reflection: Turning experience into
  learning}}}.
\newblock \bibinfo{publisher}{Kogan Page}, \bibinfo{address}{London}.
\newblock


\bibitem[\protect\citeauthoryear{Brookfield}{Brookfield}{1995}]%
        {brookfield}
\bibfield{author}{\bibinfo{person}{Stephen~D. Brookfield}.}
  \bibinfo{year}{1995}\natexlab{}.
\newblock \bibinfo{booktitle}{\emph{Becoming a critically reflective teacher}}.
\newblock \bibinfo{publisher}{Jossey-Bass}, \bibinfo{address}{San Francisco}.
\newblock


\bibitem[\protect\citeauthoryear{Coffey}{Coffey}{2017}]%
        {coffey}
\bibfield{author}{\bibinfo{person}{John~W. Coffey}.}
  \bibinfo{year}{2017}\natexlab{}.
\newblock \showarticletitle{{A Study of the Use of a Reflective Activity to
  Improve Students' Software Design Capabilities}}. In
  \bibinfo{booktitle}{\emph{Proceedings of the 2017 ACM SIGCSE Technical
  Symposium on Computer Science Education}}. Association for Computing
  Machinery, \bibinfo{publisher}{Association for Computing Machinery},
  \bibinfo{address}{Brisbane, Queensland, Australia},
  \bibinfo{pages}{129--134}.
\newblock
\urldef\tempurl%
\url{https://doi.org/10.1145/3017680.3017770}
\showDOI{\tempurl}


\bibitem[\protect\citeauthoryear{DeChenne, Koziol, Needham, and
  Enochs}{DeChenne et~al\mbox{.}}{2015}]%
        {dechenne}
\bibfield{author}{\bibinfo{person}{Sue~Ellen DeChenne},
  \bibinfo{person}{Natalie Koziol}, \bibinfo{person}{Mark Needham}, {and}
  \bibinfo{person}{Larry Enochs}.} \bibinfo{year}{2015}\natexlab{}.
\newblock \showarticletitle{Modeling sources of teaching self-efficacy for
  science, technology, engineering, and mathematics graduate teaching
  assistants}.
\newblock \bibinfo{journal}{\emph{CBE—Life Sciences Education}}
  \bibinfo{volume}{14}, \bibinfo{number}{3} (\bibinfo{year}{2015}),
  \bibinfo{pages}{ar32}.
\newblock
\urldef\tempurl%
\url{https://doi.org/10.1187/cbe.14-09-0153}
\showDOI{\tempurl}


\bibitem[\protect\citeauthoryear{Dewey}{Dewey}{1933}]%
        {dewey}
\bibfield{author}{\bibinfo{person}{John Dewey}.}
  \bibinfo{year}{1933}\natexlab{}.
\newblock \showarticletitle{{How we think: A restatement of the relation of
  reflective thinking to the educational process}}.
\newblock \bibinfo{journal}{\emph{Lexington, MA: Heath}} \bibinfo{volume}{35},
  \bibinfo{number}{64} (\bibinfo{year}{1933}), \bibinfo{pages}{690--698}.
\newblock


\bibitem[\protect\citeauthoryear{Dickson}{Dickson}{2011}]%
        {Dickson}
\bibfield{author}{\bibinfo{person}{Paul~E. Dickson}.}
  \bibinfo{year}{2011}\natexlab{}.
\newblock \showarticletitle{{Using Undergraduate Teaching Assistants in a Small
  College Environment}}. In \bibinfo{booktitle}{\emph{Proceedings of the 42nd
  ACM Technical Symposium on Computer Science Education}}
  \emph{(\bibinfo{series}{SIGCSE '11})}. \bibinfo{publisher}{ACM},
  \bibinfo{address}{New York, NY, USA}, \bibinfo{pages}{75--80}.
\newblock
\showISBNx{978-1-4503-0500-6}
\urldef\tempurl%
\url{https://doi.org/10.1145/1953163.1953187}
\showDOI{\tempurl}


\bibitem[\protect\citeauthoryear{Dyment and O'Connell}{Dyment and
  O'Connell}{2011}]%
        {dyment}
\bibfield{author}{\bibinfo{person}{Janet~E. Dyment} {and}
  \bibinfo{person}{Timothy~S. O'Connell}.} \bibinfo{year}{2011}\natexlab{}.
\newblock \showarticletitle{Assessing the quality of reflection in student
  journals: A review of the research}.
\newblock \bibinfo{journal}{\emph{Teaching in Higher Education}}
  \bibinfo{volume}{16}, \bibinfo{number}{1} (\bibinfo{year}{2011}),
  \bibinfo{pages}{81--97}.
\newblock
\urldef\tempurl%
\url{https://doi.org/10.1080/13562517.2010.507308}
\showDOI{\tempurl}


\bibitem[\protect\citeauthoryear{Epp}{Epp}{2008}]%
        {epp}
\bibfield{author}{\bibinfo{person}{Sheila Epp}.}
  \bibinfo{year}{2008}\natexlab{}.
\newblock \showarticletitle{The value of reflective journaling in undergraduate
  nursing education: A literature review}.
\newblock \bibinfo{journal}{\emph{International journal of nursing studies}}
  \bibinfo{volume}{45}, \bibinfo{number}{9} (\bibinfo{year}{2008}),
  \bibinfo{pages}{1379--1388}.
\newblock
\urldef\tempurl%
\url{https://doi.org/10.1016/j.ijnurstu.2008.01.006}
\showDOI{\tempurl}


\bibitem[\protect\citeauthoryear{Estrada and Tafliovich}{Estrada and
  Tafliovich}{2017}]%
        {Estrada}
\bibfield{author}{\bibinfo{person}{Francisco~J. Estrada} {and}
  \bibinfo{person}{Anya Tafliovich}.} \bibinfo{year}{2017}\natexlab{}.
\newblock \showarticletitle{{Bridging the Gap Between Desired and Actual
  Qualifications of Teaching Assistants: An Experience Report}}. In
  \bibinfo{booktitle}{\emph{Proceedings of the 2017 ACM Conference on
  Innovation and Technology in Computer Science Education}}
  \emph{(\bibinfo{series}{ITiCSE '17})}. \bibinfo{publisher}{ACM},
  \bibinfo{address}{New York, NY, USA}, \bibinfo{pages}{134--139}.
\newblock
\showISBNx{978-1-4503-4704-4}
\urldef\tempurl%
\url{https://doi.org/10.1145/3059009.3059023}
\showDOI{\tempurl}


\bibitem[\protect\citeauthoryear{Fatemipour}{Fatemipour}{2013}]%
        {Fatemi}
\bibfield{author}{\bibinfo{person}{Hamidreza Fatemipour}.}
  \bibinfo{year}{2013}\natexlab{}.
\newblock \showarticletitle{{The Efficiency of the Tools Used for Reflective
  Teaching in {ESL} Contexts}}.
\newblock \bibinfo{journal}{\emph{Procedia - Social and Behavioral Sciences}}
  \bibinfo{volume}{93} (\bibinfo{date}{oct} \bibinfo{year}{2013}),
  \bibinfo{pages}{1398--1403}.
\newblock
\urldef\tempurl%
\url{https://doi.org/10.1016/j.sbspro.2013.10.051}
\showDOI{\tempurl}


\bibitem[\protect\citeauthoryear{Finlay}{Finlay}{2008}]%
        {finl}
\bibfield{author}{\bibinfo{person}{Linda Finlay}.}
  \bibinfo{year}{2008}\natexlab{}.
\newblock \showarticletitle{Reflecting on reflective practice}.
\newblock \bibinfo{journal}{\emph{PBPL paper}}  \bibinfo{volume}{52}
  (\bibinfo{year}{2008}), \bibinfo{pages}{1--27}.
\newblock


\bibitem[\protect\citeauthoryear{Forbes, Malan, Pon-Barry, Reges, and
  Sahami}{Forbes et~al\mbox{.}}{2017}]%
        {Forbes}
\bibfield{author}{\bibinfo{person}{Jeffrey Forbes}, \bibinfo{person}{David~J.
  Malan}, \bibinfo{person}{Heather Pon-Barry}, \bibinfo{person}{Stuart Reges},
  {and} \bibinfo{person}{Mehran Sahami}.} \bibinfo{year}{2017}\natexlab{}.
\newblock \showarticletitle{{Scaling Introductory Courses Using Undergraduate
  Teaching Assistants}}. In \bibinfo{booktitle}{\emph{Proceedings of the 2017
  ACM SIGCSE Technical Symposium on Computer Science Education}}
  \emph{(\bibinfo{series}{SIGCSE '17})}. \bibinfo{publisher}{ACM},
  \bibinfo{address}{New York, NY, USA}, \bibinfo{pages}{657--658}.
\newblock
\showISBNx{978-1-4503-4698-6}
\urldef\tempurl%
\url{https://doi.org/10.1145/3017680.3017694}
\showDOI{\tempurl}


\bibitem[\protect\citeauthoryear{Gibbs and Coffey}{Gibbs and Coffey}{2004}]%
        {Gibbs2004}
\bibfield{author}{\bibinfo{person}{Graham Gibbs} {and} \bibinfo{person}{Martin
  Coffey}.} \bibinfo{year}{2004}\natexlab{}.
\newblock \showarticletitle{{The Impact of Training of University Teachers on
  their Teaching Skills, their Approach to Teaching and the Approach to
  Learning of their Students}}.
\newblock \bibinfo{journal}{\emph{Active Learning in Higher Education}}
  \bibinfo{volume}{5}, \bibinfo{number}{1} (\bibinfo{date}{mar}
  \bibinfo{year}{2004}), \bibinfo{pages}{87--100}.
\newblock
\urldef\tempurl%
\url{https://doi.org/10.1177/1469787404040463}
\showDOI{\tempurl}


\bibitem[\protect\citeauthoryear{Gilbert and Gibbs}{Gilbert and Gibbs}{1999}]%
        {Gilbert}
\bibfield{author}{\bibinfo{person}{Annette Gilbert} {and}
  \bibinfo{person}{Graham Gibbs}.} \bibinfo{year}{1999}\natexlab{}.
\newblock \showarticletitle{{A Proposal for an International Collaborative
  Research Programme to Identify the Impact of Initial Training on University
  Teachers}}. In \bibinfo{booktitle}{\emph{Research and Development in Higher
  Education}}. \bibinfo{publisher}{HERDSA}, \bibinfo{address}{Hammondville},
  \bibinfo{pages}{131--143}.
\newblock


\bibitem[\protect\citeauthoryear{Harvey, Coulson, and McMaugh}{Harvey
  et~al\mbox{.}}{2016}]%
        {harvey}
\bibfield{author}{\bibinfo{person}{Marina Harvey}, \bibinfo{person}{Debra
  Coulson}, {and} \bibinfo{person}{Anne McMaugh}.}
  \bibinfo{year}{2016}\natexlab{}.
\newblock \showarticletitle{{Towards a theory of the Ecology of Reflection:
  Reflective practice for experiential learning in higher education}}.
\newblock \bibinfo{journal}{\emph{Journal of University Teaching \& Learning
  Practice}} \bibinfo{volume}{13}, \bibinfo{number}{2} (\bibinfo{year}{2016}),
  \bibinfo{pages}{2}.
\newblock


\bibitem[\protect\citeauthoryear{Hatton and Smith}{Hatton and Smith}{1995}]%
        {Hatton}
\bibfield{author}{\bibinfo{person}{Neville Hatton} {and} \bibinfo{person}{David
  Smith}.} \bibinfo{year}{1995}\natexlab{}.
\newblock \showarticletitle{{Reflection in teacher education: Towards
  definition and implementation}}.
\newblock \bibinfo{journal}{\emph{Teaching and Teacher Education}}
  \bibinfo{volume}{11}, \bibinfo{number}{1} (\bibinfo{year}{1995}),
  \bibinfo{pages}{33--49}.
\newblock
\urldef\tempurl%
\url{https://doi.org/10.1016/0742-051x(94)00012-u}
\showDOI{\tempurl}


\bibitem[\protect\citeauthoryear{Hussein}{Hussein}{2006}]%
        {hussein}
\bibfield{author}{\bibinfo{person}{Jeylan~Wolyie Hussein}.}
  \bibinfo{year}{2006}\natexlab{}.
\newblock \showarticletitle{{Which One Is Better: Saying Student Teachers Don't
  Reflect or Systematically Unlocking Their Reflective Potentials: A Positive
  Experience from a Poor Teacher Education Faculty in Ethiopia}}.
\newblock \bibinfo{journal}{\emph{Australian Journal of Teacher Education}}
  \bibinfo{volume}{31}, \bibinfo{number}{2} (\bibinfo{year}{2006}),
  \bibinfo{pages}{1}.
\newblock


\bibitem[\protect\citeauthoryear{Jiang and Hill}{Jiang and Hill}{2018}]%
        {Heng}
\bibfield{editor}{\bibinfo{person}{Heng Jiang} {and} \bibinfo{person}{Mary~F.
  Hill}} (Eds.). \bibinfo{year}{2018}\natexlab{}.
\newblock \bibinfo{booktitle}{\emph{{Teacher Learning with Classroom
  Assessment}}}.
\newblock \bibinfo{publisher}{Springer}, \bibinfo{address}{Singapore}.
\newblock
\urldef\tempurl%
\url{https://doi.org/10.1007/978-981-10-9053-0}
\showDOI{\tempurl}


\bibitem[\protect\citeauthoryear{Kolb}{Kolb}{1984}]%
        {kolb}
\bibfield{author}{\bibinfo{person}{David Kolb}.}
  \bibinfo{year}{1984}\natexlab{}.
\newblock \bibinfo{booktitle}{\emph{Experiential learning as the science of
  learning and development}}.
\newblock \bibinfo{publisher}{Prentice Hall}, \bibinfo{address}{Englewood
  Cliffs, NJ}.
\newblock


\bibitem[\protect\citeauthoryear{Krathwohl}{Krathwohl}{2002}]%
        {bloom}
\bibfield{author}{\bibinfo{person}{David~R. Krathwohl}.}
  \bibinfo{year}{2002}\natexlab{}.
\newblock \showarticletitle{{A revision of Bloom's taxonomy: An overview}}.
\newblock \bibinfo{journal}{\emph{Theory into practice}} \bibinfo{volume}{41},
  \bibinfo{number}{4} (\bibinfo{year}{2002}), \bibinfo{pages}{212--218}.
\newblock
\urldef\tempurl%
\url{https://doi.org/10.1207/s15430421tip4104_2}
\showDOI{\tempurl}


\bibitem[\protect\citeauthoryear{Leyzberg, Lumbroso, and Moretti}{Leyzberg
  et~al\mbox{.}}{2017}]%
        {Leyzberg}
\bibfield{author}{\bibinfo{person}{Dan Leyzberg},
  \bibinfo{person}{J{\'e}r{\'e}mie Lumbroso}, {and}
  \bibinfo{person}{Christopher Moretti}.} \bibinfo{year}{2017}\natexlab{}.
\newblock \showarticletitle{{Nailing the TA Interview: Using a Rubric to Hire
  Teaching Assistants}}. In \bibinfo{booktitle}{\emph{Proceedings of the 2017
  ACM Conference on Innovation and Technology in Computer Science Education}}
  \emph{(\bibinfo{series}{ITiCSE '17})}. \bibinfo{publisher}{ACM},
  \bibinfo{address}{New York, NY, USA}, \bibinfo{pages}{128--133}.
\newblock
\urldef\tempurl%
\url{https://doi.org/10.1145/3059009.3059057}
\showDOI{\tempurl}


\bibitem[\protect\citeauthoryear{Lindroth}{Lindroth}{2015}]%
        {lindroth}
\bibfield{author}{\bibinfo{person}{James~T. Lindroth}.}
  \bibinfo{year}{2015}\natexlab{}.
\newblock \showarticletitle{{Reflective journals: A review of the literature}}.
\newblock \bibinfo{journal}{\emph{Update: Applications of Research in Music
  Education}} \bibinfo{volume}{34}, \bibinfo{number}{1} (\bibinfo{year}{2015}),
  \bibinfo{pages}{66--72}.
\newblock
\urldef\tempurl%
\url{https://doi.org/10.1177/8755123314548046}
\showDOI{\tempurl}


\bibitem[\protect\citeauthoryear{Loughran}{Loughran}{2002}]%
        {loug}
\bibfield{author}{\bibinfo{person}{John Loughran}.}
  \bibinfo{year}{2002}\natexlab{}.
\newblock \showarticletitle{{Effective reflective practice: In search of
  meaning in learning about teaching}}.
\newblock \bibinfo{journal}{\emph{Journal of teacher education}}
  \bibinfo{volume}{53}, \bibinfo{number}{1} (\bibinfo{year}{2002}),
  \bibinfo{pages}{33--43}.
\newblock
\urldef\tempurl%
\url{https://doi.org/10.1177/0022487102053001004}
\showDOI{\tempurl}


\bibitem[\protect\citeauthoryear{{Martin Ukrop, Ondřej Přibyla, Valdemar
  Švábenský, and Martin Macák}}{{Martin Ukrop, Ondřej Přibyla, Valdemar
  Švábenský, and Martin Macák}}{2017}]%
        {diary}
\bibfield{author}{\bibinfo{person}{{Martin Ukrop, Ondřej Přibyla, Valdemar
  Švábenský, and Martin Macák}}.} \bibinfo{year}{2017}\natexlab{}.
\newblock \bibinfo{title}{{Teacher's Reflective Diary}}.
\newblock
\newblock
\urldef\tempurl%
\url{https://github.com/teaching-lab/reflective-diary}
\showURL{%
\tempurl}


\bibitem[\protect\citeauthoryear{McDermott, Brindley, and Eccleston}{McDermott
  et~al\mbox{.}}{2010}]%
        {mcdermott}
\bibfield{author}{\bibinfo{person}{Roger McDermott}, \bibinfo{person}{Garry
  Brindley}, {and} \bibinfo{person}{Gordon Eccleston}.}
  \bibinfo{year}{2010}\natexlab{}.
\newblock \showarticletitle{Developing tools to encourage reflection in first
  year students blogs}. In \bibinfo{booktitle}{\emph{Proceedings of the
  fifteenth annual conference on Innovation and technology in computer science
  education}}. ACM, \bibinfo{publisher}{Association for Computing Machinery},
  \bibinfo{address}{Bilkent, Ankara, Turkey}, \bibinfo{pages}{147--151}.
\newblock
\urldef\tempurl%
\url{https://doi.org/10.1145/1822090.1822132}
\showDOI{\tempurl}


\bibitem[\protect\citeauthoryear{Moon}{Moon}{2006}]%
        {moon}
\bibfield{author}{\bibinfo{person}{Jennifer~A. Moon}.}
  \bibinfo{year}{2006}\natexlab{}.
\newblock \bibinfo{booktitle}{\emph{{Learning journals: A handbook for
  reflective practice and professional development}}}.
\newblock \bibinfo{publisher}{Routledge}, \bibinfo{address}{London}.
\newblock
\urldef\tempurl%
\url{https://doi.org/10.4324/9780429448836-8}
\showDOI{\tempurl}


\bibitem[\protect\citeauthoryear{Patitsas}{Patitsas}{2013}]%
        {Patitsas}
\bibfield{author}{\bibinfo{person}{Elizabeth Patitsas}.}
  \bibinfo{year}{2013}\natexlab{}.
\newblock \showarticletitle{{A Case Study of the Development of CS Teaching
  Assistants and Their Experiences with Team Teaching}}. In
  \bibinfo{booktitle}{\emph{Proceedings of the 13th Koli Calling International
  Conference on Computing Education Research}} \emph{(\bibinfo{series}{Koli
  Calling '13})}. \bibinfo{publisher}{ACM}, \bibinfo{address}{New York, NY,
  USA}, \bibinfo{pages}{115--124}.
\newblock
\showISBNx{978-1-4503-2482-3}
\urldef\tempurl%
\url{https://doi.org/10.1145/2526968.2526981}
\showDOI{\tempurl}


\bibitem[\protect\citeauthoryear{Pedro}{Pedro}{2005}]%
        {pedro}
\bibfield{author}{\bibinfo{person}{Joan~Y. Pedro}.}
  \bibinfo{year}{2005}\natexlab{}.
\newblock \showarticletitle{Reflection in teacher education: exploring
  pre-service teachers’ meanings of reflective practice}.
\newblock \bibinfo{journal}{\emph{Reflective practice}} \bibinfo{volume}{6},
  \bibinfo{number}{1} (\bibinfo{year}{2005}), \bibinfo{pages}{49--66}.
\newblock
\urldef\tempurl%
\url{https://doi.org/10.1080/1462394042000326860}
\showDOI{\tempurl}


\bibitem[\protect\citeauthoryear{Petty}{Petty}{2009}]%
        {petty}
\bibfield{author}{\bibinfo{person}{Geoffrey Petty}.}
  \bibinfo{year}{2009}\natexlab{}.
\newblock \bibinfo{booktitle}{\emph{{Teaching Today: A Practical Guide}}}.
\newblock \bibinfo{publisher}{Nelson Thornes}, \bibinfo{address}{Cheltenham,
  UK}.
\newblock
\showISBNx{9781408504154}
\showLCCN{2010290320}


\bibitem[\protect\citeauthoryear{Reding and Dorn}{Reding and Dorn}{2017}]%
        {reding}
\bibfield{author}{\bibinfo{person}{Tracie~E. Reding} {and}
  \bibinfo{person}{Brian Dorn}.} \bibinfo{year}{2017}\natexlab{}.
\newblock \showarticletitle{{Understanding the "Teacher Experience" in Primary
  and Secondary {CS} Professional Development}}. In
  \bibinfo{booktitle}{\emph{Proceedings of the 2017 {ACM} Conference on
  International Computing Education Research - {ICER} {\textquotesingle}17}}.
  \bibinfo{publisher}{Association for Computing Machinery},
  \bibinfo{address}{Tacoma, WA}, \bibinfo{pages}{155--163}.
\newblock
\urldef\tempurl%
\url{https://doi.org/10.1145/3105726.3106185}
\showDOI{\tempurl}


\bibitem[\protect\citeauthoryear{Reges}{Reges}{2003}]%
        {Reges}
\bibfield{author}{\bibinfo{person}{Stuart Reges}.}
  \bibinfo{year}{2003}\natexlab{}.
\newblock \showarticletitle{{Using Undergraduates as Teaching Assistants at a
  State University}}. In \bibinfo{booktitle}{\emph{Proceedings of the 34th
  SIGCSE Technical Symposium on Computer Science Education}}
  \emph{(\bibinfo{series}{SIGCSE '03})}. \bibinfo{publisher}{ACM},
  \bibinfo{address}{New York, NY, USA}, \bibinfo{pages}{103--107}.
\newblock
\showISBNx{1-58113-648-X}
\urldef\tempurl%
\url{https://doi.org/10.1145/611892.611943}
\showDOI{\tempurl}


\bibitem[\protect\citeauthoryear{Roberts, Lilly, and Rollins}{Roberts
  et~al\mbox{.}}{1995}]%
        {Roberts}
\bibfield{author}{\bibinfo{person}{Eric Roberts}, \bibinfo{person}{John Lilly},
  {and} \bibinfo{person}{Bryan Rollins}.} \bibinfo{year}{1995}\natexlab{}.
\newblock \showarticletitle{{Using Undergraduates as Teaching Assistants in
  Introductory Programming Courses: An Update on the Stanford Experience}}. In
  \bibinfo{booktitle}{\emph{Proceedings of the Twenty-sixth SIGCSE Technical
  Symposium on Computer Science Education}} \emph{(\bibinfo{series}{SIGCSE
  '95})}. \bibinfo{publisher}{ACM}, \bibinfo{address}{New York, NY, USA},
  \bibinfo{pages}{48--52}.
\newblock
\showISBNx{0-89791-693-X}
\urldef\tempurl%
\url{https://doi.org/10.1145/199688.199716}
\showDOI{\tempurl}


\bibitem[\protect\citeauthoryear{{Royal College of Nursing}}{{Royal College of
  Nursing}}{2015}]%
        {diary1}
\bibfield{author}{\bibinfo{person}{{Royal College of Nursing}}.}
  \bibinfo{year}{2015}\natexlab{}.
\newblock \bibinfo{title}{{Reflective Diary}}.
\newblock
\newblock
\urldef\tempurl%
\url{http://rcneolnutritionhydration.org.uk/wp-content/uploads/sites/6/2015/11/reflective-diary-V3.pdf}
\showURL{%
\tempurl}


\bibitem[\protect\citeauthoryear{Sch{\"o}n}{Sch{\"o}n}{1983}]%
        {schon}
\bibfield{author}{\bibinfo{person}{Donald Sch{\"o}n}.}
  \bibinfo{year}{1983}\natexlab{}.
\newblock \bibinfo{booktitle}{\emph{{The reflective practitioner: How
  practitioners think in action}}}.
\newblock \bibinfo{publisher}{Temple Smith}, \bibinfo{address}{London}.
\newblock


\bibitem[\protect\citeauthoryear{{Scottish Social Services Council}}{{Scottish
  Social Services Council}}{2011}]%
        {diary3}
\bibfield{author}{\bibinfo{person}{{Scottish Social Services Council}}.}
  \bibinfo{year}{2011}\natexlab{}.
\newblock \bibinfo{title}{{Reflective Diary}}.
\newblock
\newblock
\urldef\tempurl%
\url{http://workforcesolutions.sssc.uk.com/PracticeSim/scenario_02/Diary/Reflective\%20Diary.pdf}
\showURL{%
\tempurl}


\bibitem[\protect\citeauthoryear{Siebert and Walsh}{Siebert and Walsh}{2013}]%
        {siebert}
\bibfield{author}{\bibinfo{person}{Sabina Siebert} {and} \bibinfo{person}{Anita
  Walsh}.} \bibinfo{year}{2013}\natexlab{}.
\newblock \showarticletitle{{Reflection in work-based learning: Self-regulation
  or self-liberation?}}
\newblock \bibinfo{journal}{\emph{Teaching in Higher Education}}
  \bibinfo{volume}{18}, \bibinfo{number}{2} (\bibinfo{year}{2013}),
  \bibinfo{pages}{167--178}.
\newblock
\urldef\tempurl%
\url{https://doi.org/10.1080/13562517.2012.696539}
\showDOI{\tempurl}


\bibitem[\protect\citeauthoryear{Stone}{Stone}{2012}]%
        {stone}
\bibfield{author}{\bibinfo{person}{Jeffrey~A. Stone}.}
  \bibinfo{year}{2012}\natexlab{}.
\newblock \showarticletitle{Using reflective blogs for pedagogical feedback in
  CS1}. In \bibinfo{booktitle}{\emph{Proceedings of the 43rd ACM technical
  symposium on Computer Science Education}}. ACM,
  \bibinfo{publisher}{Association for Computing Machinery},
  \bibinfo{address}{Raleigh, NC, USA}, \bibinfo{pages}{259--264}.
\newblock
\urldef\tempurl%
\url{https://doi.org/10.1145/2157136.2157216}
\showDOI{\tempurl}


\bibitem[\protect\citeauthoryear{Tembrioti and Tsangaridou}{Tembrioti and
  Tsangaridou}{2013}]%
        {Tembri}
\bibfield{author}{\bibinfo{person}{Lara Tembrioti} {and} \bibinfo{person}{Niki
  Tsangaridou}.} \bibinfo{year}{2013}\natexlab{}.
\newblock \showarticletitle{Reflective practice in dance: a review of the
  literature}.
\newblock \bibinfo{journal}{\emph{Research in Dance Education}}
  \bibinfo{volume}{15}, \bibinfo{number}{1} (\bibinfo{date}{jul}
  \bibinfo{year}{2013}), \bibinfo{pages}{4--22}.
\newblock
\urldef\tempurl%
\url{https://doi.org/10.1080/14647893.2013.809521}
\showDOI{\tempurl}


\bibitem[\protect\citeauthoryear{Topping}{Topping}{1996}]%
        {topping}
\bibfield{author}{\bibinfo{person}{Keith~J. Topping}.}
  \bibinfo{year}{1996}\natexlab{}.
\newblock \showarticletitle{The effectiveness of peer tutoring in further and
  higher education: A typology and review of the literature}.
\newblock \bibinfo{journal}{\emph{Higher Education}} \bibinfo{volume}{32},
  \bibinfo{number}{3} (\bibinfo{date}{oct} \bibinfo{year}{1996}),
  \bibinfo{pages}{321--345}.
\newblock
\urldef\tempurl%
\url{https://doi.org/10.1007/bf00138870}
\showDOI{\tempurl}


\bibitem[\protect\citeauthoryear{Trif and Popescu}{Trif and Popescu}{2013}]%
        {trif}
\bibfield{author}{\bibinfo{person}{Letitia Trif} {and} \bibinfo{person}{Teodora
  Popescu}.} \bibinfo{year}{2013}\natexlab{}.
\newblock \showarticletitle{The reflective diary, an effective professional
  training instrument for future teachers}.
\newblock \bibinfo{journal}{\emph{Procedia-Social and Behavioral Sciences}}
  \bibinfo{volume}{93} (\bibinfo{year}{2013}), \bibinfo{pages}{1070--1074}.
\newblock
\urldef\tempurl%
\url{https://doi.org/10.1016/j.sbspro.2013.09.332}
\showDOI{\tempurl}


\bibitem[\protect\citeauthoryear{Vihavainen, Vikberg, Luukkainen, and
  Kurhila}{Vihavainen et~al\mbox{.}}{2013}]%
        {Vihavainen}
\bibfield{author}{\bibinfo{person}{Arto Vihavainen}, \bibinfo{person}{Thomas
  Vikberg}, \bibinfo{person}{Matti Luukkainen}, {and} \bibinfo{person}{Jaakko
  Kurhila}.} \bibinfo{year}{2013}\natexlab{}.
\newblock \showarticletitle{{Massive Increase in Eager TAs: Experiences from
  Extreme Apprenticeship-based CS1}}. In \bibinfo{booktitle}{\emph{Proceedings
  of the 18th ACM Conference on Innovation and Technology in Computer Science
  Education}} \emph{(\bibinfo{series}{ITiCSE '13})}. \bibinfo{publisher}{ACM},
  \bibinfo{address}{New York, NY, USA}, \bibinfo{pages}{123--128}.
\newblock
\urldef\tempurl%
\url{https://doi.org/10.1145/2462476.2462508}
\showDOI{\tempurl}


\bibitem[\protect\citeauthoryear{Wallin and Adawi}{Wallin and Adawi}{2018}]%
        {wallin}
\bibfield{author}{\bibinfo{person}{Patric Wallin} {and} \bibinfo{person}{Tom
  Adawi}.} \bibinfo{year}{2018}\natexlab{}.
\newblock \showarticletitle{The reflective diary as a method for the formative
  assessment of self-regulated learning}.
\newblock \bibinfo{journal}{\emph{European Journal of Engineering Education}}
  \bibinfo{volume}{43}, \bibinfo{number}{4} (\bibinfo{year}{2018}),
  \bibinfo{pages}{507--521}.
\newblock
\urldef\tempurl%
\url{https://doi.org/10.1080/03043797.2017.1290585}
\showDOI{\tempurl}


\end{thebibliography}

\end{document}